\documentclass[conference]{IEEEtran}

\usepackage{amsmath}
\usepackage{tabularx}
\usepackage{graphicx}
\usepackage{amssymb}
\usepackage{amsthm}
\usepackage{color}
\usepackage{cite}
\usepackage{algorithm,algpseudocode}
\usepackage{tikz}
\usetikzlibrary{arrows.meta,positioning,fit,backgrounds,shapes.geometric}
\usepackage{url}

\usepackage{enumitem}

\usepackage[top=0.75in, bottom=1in, left=0.625in, right=0.625in]{geometry}

\tikzset{
  blk/.style  ={draw, rounded corners=3pt, fill=white,
                 minimum width=1.7cm, minimum height=0.6cm,
                 align=center, font=\footnotesize},
  gemblk/.style={blk, fill=blue!8, draw=blue!50!black},
  ppobl/.style ={blk, fill=orange!10, draw=orange!60!black},
  envbl/.style ={blk, fill=gray!12,  draw=gray!60},
  bagbl/.style ={draw, cylinder, fill=green!8, draw=green!50!black,
                 minimum width=1.8cm, minimum height=0.8cm,
                 align=center, font=\footnotesize, shape border rotate=90, aspect=0.25},
  embbl/.style ={blk, fill=purple!8, draw=purple!50!black},
  rtbl/.style  ={blk, fill=yellow!10,draw=black!60},
  arr/.style   ={-{Stealth[length=4pt]}, thick},
  darr/.style  ={-{Stealth[length=4pt]}, thick, dashed},
  lbl/.style   ={font=\scriptsize, inner sep=1pt, fill=white, fill opacity=0.8, text opacity=1},
}

\IEEEoverridecommandlockouts

\usepackage{fancyhdr}
\fancypagestyle{preprintfirstpage}{%
  \fancyhf{}%
  \fancyfoot[C]{%
    \parbox[b]{\textwidth}{%
      \centering\fontsize{7}{8}\selectfont
      \textit{Accepted for publication at IEEE GLOBECOM. Proceedings forthcoming.}\par
      \textcopyright~2026 IEEE. Personal use of this material is permitted. Permission from IEEE must be obtained for all other uses, in any current or future media, including reprinting/republishing this material for advertising or promotional purposes, creating new collective works, for resale or redistribution to servers or lists, or reuse of any copyrighted component of this work in other works.%
    }%
  }%
}

\begin{document}

\title{\textit{Ask the Expert}: LLM-Guided Reinforcement Learning for Autonomous Cyber Defense}

\author{
\IEEEauthorblockN{Fernando Martinez$^\dagger$, Abhishek Satyam$^\dagger$, Tao Li, Junaid Farooq, Ying Wang, and Juntao Chen
}
\thanks{$^\dagger$Equal contribution.}
\thanks{F. Martinez, A. Satyam, and J. Chen are with the Department of Computer and Information Sciences, Fordham University, New York, USA. \{fmartinezlopez, as526, jchen504\}@fordham.edu}
\thanks{T. Li is with the Department of Systems Engineering,  City University of Hong Kong, Hong Kong SAR, China. T. Li is supported by a CityU internal grant No. 9610792.}
\thanks{J. Farooq is with the Department of Electrical and Computer Engineering, University of Michigan-Dearborn, Dearborn, MI, USA.}
\thanks{Y. Wang is with the Department of Systems Engineering, Stevens Institute of Technology, New Jersey, USA.}
}
\maketitle
\thispagestyle{preprintfirstpage}

\begin{abstract}
Policy-based reinforcement learning (RL) approaches have produced promising results for autonomous cyber defense; however, they are sample-inefficient in settings where defenders must respond under delayed, partial observations with actions from large action spaces. While large language models (LLMs) may reason semantically about security state space, high latency and trust assumptions prevent attractive in-line deployment models. We introduce \emph{Ask the Expert}, a training-time guidance framework which first summarizes hard cyber-defense states, then intermittently queries an LLM for host-level defensive recommendations via a constrained action interface, and finally transforms those recommendations into tiered reward shaping for use with PPO. Because the LLM is discarded after training, deployment is a pure RL policy. Across TTCP CAGE CC1 and CC2 and both attacker types, this asymmetric design improves sample efficiency over PPO and outperforms the evaluated potential-based reward shaping (PBRS) baselines, while retaining the strongest terminal mean and requiring no LLM dependency at deployment time.
\end{abstract}


\section{Introduction}
\label{sec:introduction}

Enterprise networks are subject to continual risk from post-compromise lateral movement campaigns that slowly infiltrate their networks by reusing credentials and pivoting across hosts~\cite{smiliotopoulos2024detecting}. Embedded autonomous agents intended to respond to these intrusions must act under tight time constraints and with incomplete network observability. Successfully responding requires agents to reason over local alerts and choose from a large, semantically dense action space (e.g., conducting host forensics, terminating sessions, and deploying remediation), while being predominantly guided by dense but tactically weak reward signals.

Reinforcement learning (RL), and policy gradient methods like PPO~\cite{schulman2017proximal} in particular, scale to learning policies over large structured discrete action spaces~\cite{nguyen2019deep,hu2021survey}. Unfortunately, applying RL solutions directly to the response problem can behave destructively during early exploration: unfettered RL agents will drive significant service degradation for normal users until eventually stumbling onto reasonable behavior~\cite{bates2023reward}. Moreover, the scalar reward emitted by the environment represents global network health, but does not provide meaningful, tactically valid feedback for granular actions an agent takes~\cite{palmer2023deep}. The bulk of early agent training time is spent performing no-knowledge exploration until converging on a reasonable policy, if at all. Classical potential-based reward shaping (PBRS) provides the theoretical backdrop for altering rewards while preserving optimal policies under restricted transformations~\cite{ng1999_rewardshaping}, but practical cyber defense still relies heavily on heuristic shaping rules. We address this gap by using an empirically inspired, training-time shaping scheme derived from LLM advice rather than claiming policy invariance.

Separately, Large Language Models (LLMs) have been applied to qualitative cybersecurity reasoning tasks~\cite{xu2025llm4sec,purves2024causally,zhang2023large} as well as reward generation for automated RL problems~\cite{ma2024eureka, pmlr-v202-du23f}. But interacting with an LLM directly as an online runtime agent would incur unacceptable inference latency and present dangerous trust-boundary risks. Fortunately, we can leverage similar asymmetric training regimes from SCORER~\cite{martinez2026stackelberg} and CORAL~\cite{martinez2025context} that allow us to inject enriched auxiliary knowledge during training-time. This leaves us with the question: how can we harness the zero-shot reasoning power of a pre-trained LLM as an asymmetric teacher?

We present \emph{Ask the Expert}, an LLM-guided RL framework that intermittently samples hard states in which the defender first observes a newly spreading compromise. These states are passed to a zero-shot LLM, which returns host-focused remediation advice. The advice is distilled into a persistent tiered reward-shaping policy used only during training. After training, the LLM is discarded, leaving a fast and fully autonomous deployment policy.

Our contributions are as follows:

\begin{itemize}[leftmargin=*,topsep=2pt,itemsep=1pt]

\item We introduce \textbf{Ask the Expert}, a training-time-only LLM-guided reward-shaping framework for autonomous cyber defense that improves RL learning without requiring any deployment-time LLM calls or added inference latency.

\item We design a \textbf{recent-frontier prompting mechanism} that stores hard states involving newly spreading compromises, prioritizes recent severe failures, caps prompt concentration on any one spreading host, and distills host-level LLM guidance into persistent tiered priors over remediation actions.

\item We conduct a \textbf{multi-seed evaluation} on CAGE CC1 and CC2 against two attacker types, showing improved sample efficiency over PPO and observation-based PBRS baselines.
\end{itemize}

\section{Related Work}
\label{sec:related}

\textbf{RL for autonomous cyber defense.} Past research has explored the use of RL for broader network intrusion detection and response efforts~\cite{nguyen2019deep,hu2021survey,kim-tao25col}. These environments must balance attack containment against availability under partial observability. Existing benchmarks like CybORG helped standardize the problem~\cite{cyborg2021}. Prior solutions often relied on exploration through heuristic reward shaping or action masking~\cite{bates2023reward, palmer2023deep}. However, these schemes rely on static rules curated by domain experts. Our recent-frontier prompt sampler is closer in spirit to prioritized selection~\cite{schaul2016_per} than to uniform replay, although it prioritizes LLM prompt candidates rather than TD updates. We then convert the resulting advice into dynamically tiered reward bonuses rather than fixed hand-designed heuristics.

\textbf{Auxiliary signals and structured advice for RL.} Structured training signals can improve RL sample efficiency and generalization without deployment overhead. Policy Shaping~\cite{griffith2013_policyshaping} was among the first to translate external advice into policy guidance rather than reward relabeling. Other asymmetric-training methods use auxiliary signals to guide representations or messages, including SCORER~\cite{martinez2026stackelberg} and CORAL~\cite{martinez2025context}. Wang et al.~\cite{wang2026collaborative} combine LLM-generated action demonstrations and reward shaping with constrained PPO for cloud-edge resource allocation. In contrast, we use intermittent host-level advice only as training-time reward bonuses, without demonstrations or deployment-time LLM calls.

\textbf{LLMs for RL and Security.} Large language models have been applied to RL reward design: Eureka~\cite{ma2024eureka} and Text2Reward~\cite{xie2024textreward} generate reward code from natural-language descriptions, while ELLM~\cite{pmlr-v202-du23f} and PRM4RL~\cite{zhang2025progress} prompt for subgoals or progress rewards; CARD~\cite{sun2025_card} and LLMHRL~\cite{li2026_llmhrl} apply LLMs to reward reshaping and hierarchical control. In security, deployment latency and trust concerns make direct runtime LLM use unattractive~\cite{zhang2023large, purves2024causally, xu2025llm4sec}. In contrast, we use intermittent LLM advice only at training time, operationalized as reward bonuses, then discard the LLM at deployment.

\section{Problem Formulation}
\label{sec:problem_formulation}

\subsection{Modeling Network Defense as a POMDP}
We model autonomous intrusion response as a single-agent partially observable Markov decision process (POMDP) $\left(\mathcal{S}, \mathcal{O}, \mathcal{A}, \mathcal{T}, r, \gamma\right)$, in which an autonomous Blue defender interacts with a concurrent adversarial Red agent. The true global state of the world $s_t \in \mathcal{S}$ is unknown. Instead, the defender receives delayed, partial observations $o_t \in \mathcal{O}$. These observations are based on host alerts and network telemetry, indicative of the consequences of compromise rather than direct access to the attacker's location. At no point during training or deployment does the defender receive observations of Red's internal state or intended future plan.

\subsection{Threat Model}
We focus on adversarial post-compromise lateral movement~\cite{homoliak2019insight} in CAGE CC1 and CC2. Red starts every episode with a foothold at \texttt{User0}. We benchmark against two attacker policies: \texttt{B\_lineAgent}, which attempts a direct pivot toward the operational server, and \texttt{RedMeanderAgent}, which sweeps subnets before penetrating inward. Regardless, Blue reacts solely to defender-visible aftermath signals.

\subsection{Action Space and Semantics}
The defender operates over a large discrete verb--host action space $\mathcal{A}$. The PPO agent acts over the entire environment action space, while the LLM-facing action interface only exposes a much smaller host-level remediation subspace. This allows us to maintain a much simpler language interface while still retaining full fidelity of the learned policy.

\subsection{Reward and Objective}
The environment reward function $r_t$ penalizes compromised hosts based on their criticality to operations. Negative effects of being overly aggressive, such as unnecessarily disrupting benign services that were not threats, are captured in the resulting aggregate reward. The policy aims to maximize discounted return:
\begin{equation}
J(\pi) = \mathbb{E}_\pi\!\left[\sum_{t=0}^{T} \gamma^t r_t\right].
\label{eq:objective}
\end{equation}

Hidden penalties continuously obfuscate the tactical value of immediate actions, leading rudimentary agents either to hesitate during active pivoting or to blindly disrupt critical services or subnets. The role of our framework is therefore to add semantic training-time guidance without changing the underlying environment dynamics or deployment-time policy class.

\section{Ask the Expert Framework}
\label{sec:method}

\begin{figure*}[!t]
\centering
\resizebox{1.0\textwidth}{!}{%
\begin{tikzpicture}[>=latex]
  \node[ppobl] (agent) at (0, 0) {RL Agent\\(Policy $\pi_\theta$)};
  \node[envbl] (env) at (0, -4.5) {Network\\Environment};

  \node[rtbl] (shaping) at (4.5, 0) {Rollout-End Tiered\\Reward Shaping\\($+\beta \cdot w_i$)};
  \node[embbl] (expansion) at (10, 0) {Merged Host Policy\\$\Pi_{\text{LLM}}$};

  \node[gemblk] (llm) at (10, -2.25) {LLM (e.g., Gemini)\\Zero-Shot Expert};
  \node[blk] (prompt) at (15.5, -2.25) {Recent-Frontier Sampler\\\& Prompt Builder};

  \node[bagbl] (bag) at (15.5, -4.5) {Candidate Bag $\mathcal{C}$\\(Recent States)};
  \node[blk] (filter) at (9.5, -4.5) {Spreading-State Filter\\($|S_t| > 0$)};
  \node[blk] (encoder) at (4.5, -4.5) {Semantic Encoder $\phi$};

  \draw[arr] ([xshift=-12pt]agent.south) -- ([xshift=-12pt]env.north) node[midway, left, lbl] {Action $a_t$};
  \draw[arr] ([xshift=12pt]env.north) -- ([xshift=12pt]agent.south) node[midway, right, lbl] {Obs $o_{t+1}$};

  \draw[arr] (env.north east) -| (shaping.south) node[pos=0.58, left, lbl] {Rollout $r_t$};
  \draw[arr] (shaping.west) -- (agent.east) node[midway, above, lbl, align=center] {Shaped\\Reward $\tilde{r}_t$};

  \draw[arr] (env.east) -- (encoder.west) node[pos=0.62, below, lbl, align=center] {Reward\\Breakdown};
  \draw[arr] (encoder.east) -- (filter.west) node[midway, above, lbl] {Encoded State};
  \draw[arr] (filter.east) -- (bag.west) node[midway, above, lbl] {Push Hard State};

  \draw[arr] (bag.north) -- (prompt.south) node[midway, left, lbl] {Recent Frontier Sample};
  \draw[arr] (prompt.west) -- (llm.east) node[midway, above, lbl] {Context Prompt $p$};

  \draw[arr] (llm.north) -- (expansion.south) node[midway, right, lbl] {JSON Actions};
  \draw[arr] (expansion.west) -- (shaping.east) node[midway, above, lbl, align=center] {Tiered Action\\Weights};

  \begin{scope}[on background layer]
    \node[draw=orange!50, dashed, thick, rounded corners, fit=(agent) (env), inner sep=15pt] (rl_loop) {};
    \node[above, font=\bfseries\small, text=orange!80!black] at (rl_loop.north) {RL Interaction Loop};

    \node[draw=blue!50, dashed, thick, rounded corners, fit=(encoder) (filter) (bag) (prompt) (llm) (expansion) (shaping), inner sep=15pt] (expert_loop) {};
    \node[above, font=\bfseries\small, text=blue!80!black] at (expert_loop.north) {Ask the Expert Framework (Training-Time Only)};
  \end{scope}
\end{tikzpicture}
}
\caption{The \emph{Ask the Expert} architecture shared across both \texttt{B\_lineAgent} and \texttt{RedMeanderAgent} experiments. Reward-breakdown telemetry is encoded into compromised, spreading, and clean host sets; spreading hard states are stored in the candidate bag; a recent-frontier sampler builds the LLM prompt; and the returned host-level advice is merged into a tiered reward-shaping policy applied at rollout end before PPO advantage computation. Deployment remains pure RL with no LLM calls.}
\label{fig:framework}
\vspace{-10pt}
\end{figure*}
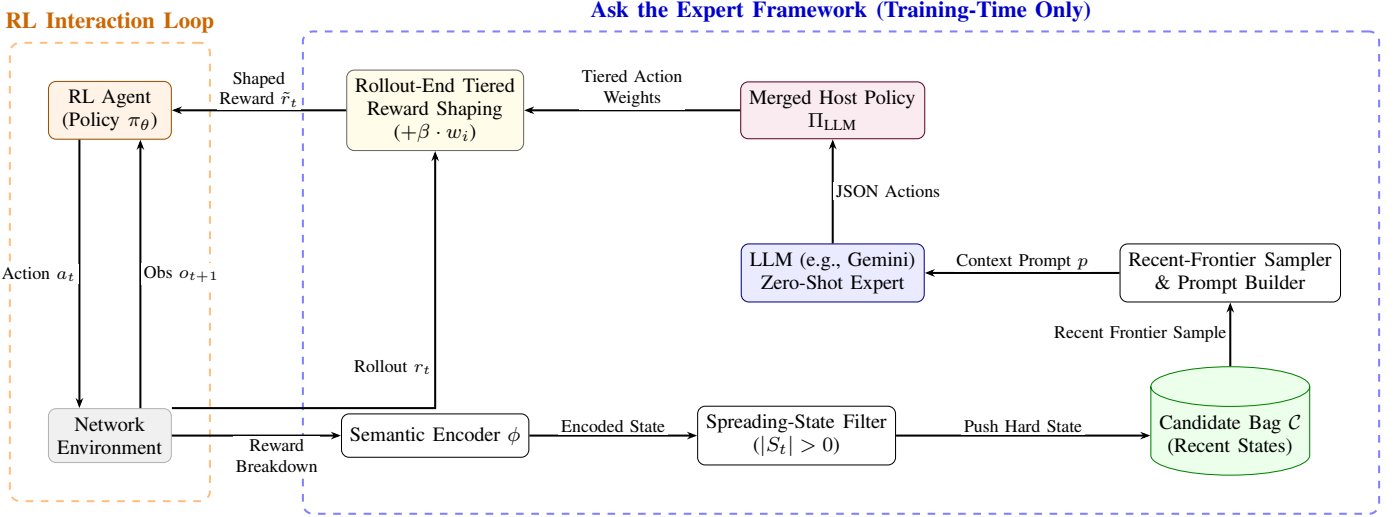

\subsection{Framework Architecture}

Our framework supplements a standard policy-gradient RL agent with three modules run at training time (Figure~\ref{fig:framework}): (i) a \emph{semantic encoder} $\phi$ that converts defender reward-breakdown telemetry into sets of compromised hosts, newly spreading hosts, and clean hosts; (ii) a fixed-capacity \emph{candidate bag} $\mathcal{C}$ and recent-frontier sampler that stores recent spreading-state transitions and periodically queries an LLM; and (iii) a \emph{tiered reward-shaping policy} $\Pi_{\text{LLM}}$ that produces host-specific bonus weights for semantically related actions. PPO always takes its own actions; the LLM never intervenes on action selection and can only affect training via reward shaping. The LLM is discarded after training, leaving only $\pi_\theta$.

\subsection{Semantic State Encoder}

The semantic encoder $\phi$ is deterministic. At each timestep, it inputs defender-visible reward-breakdown telemetry along with the previously known set of compromised hosts, and outputs three summaries: compromised hosts, newly detected hosts, and clean hosts. Newly detected hosts comprise the spreading set
\begin{equation}
S_t = \text{Compromised}_t \setminus \text{Compromised}_{t-1}.
\end{equation}

The produced semantic summary $\zeta_t$ is compact and easily digestible by a general-purpose language model. The summary exposes defender-available information and contains no test-time dependencies on the simulator's hidden state.

\subsection{Hard-State Queue and LLM Querying}
We maintain a fixed-capacity candidate bag $\mathcal{C}$ of spreading-state transitions. This bag stores $(\zeta_t, a_t, r_t, \Delta r_t, n)$ whenever $S_t \neq \emptyset$, where $\Delta r_t = r_t - r_{t-1}$ is stored as context for the LLM but is unused as a storage gate. Entries are dropped from the head of $\mathcal{C}$ whenever $|\mathcal{C}| > B_{\max}$.

Once per LLM query iteration, the sampler filters $\mathcal{C}$ down to a frontier of recent failures, ranks entries by the most negative reward deltas, caps any single spreading host at three prompt slots, and selects up to $n_s$ entries subject to that host cap. The resulting prompt $p$ conditions on both network context and a maintained history set $\mathcal{H}$ of past LLM suggestions and their observed outcomes, and for each sampled hard state appends the semantic summary, PPO's selected action, and a qualitative outcome tag indicating reward deterioration, no change, or reward improvement. The LLM is constrained to a fixed host-level remediation vocabulary at host granularity, and invalid outputs are rejected during parsing; in CC2 this vocabulary contains \texttt{Analyse}, \texttt{Remove}, and \texttt{Restore}, while in CC1 it additionally includes \texttt{Misinform}, which deploys a decoy service on a host to delay and expose attacker lateral movement. This frontier-sampled host-level guidance mechanism is shared across attacker configurations, while allowing the prompt doctrine itself to be specialized to attacker structure (e.g. \texttt{B\_lineAgent}'s direct pivots vs. \texttt{RedMeanderAgent}'s subnet sweeps).

\subsection{Tiered LLM-Guided Reward Shaping}
To eliminate deployment-time dependence on the LLM, we utilize its suggestions only through reward-shaping during training. For each host $h$, the shaping module stores a tiered policy:
\begin{equation}
\Pi_{\text{LLM}}(h) = \{(a_i, w_i)\},
\end{equation}
where $a_i$ is an action string and $w_i \in (0,1]$ is that action's bonus weight. Each host is initialized with weak default tiers over conservative remediation actions. After receiving a valid LLM response, its suggestion becomes $\Pi_{\text{LLM}}(h)$'s single high-weight primary tier, while semantically-related actions pertaining to host $h$ are inserted as fallback tiers with lower weight. Specific tier weights are determined below as part of experimental instantiation.

After rollout concludes and before generalized advantage estimation (GAE), the method revisits each buffered timestep associated with a spreading host $h_t$. If PPO chose action $a_t$ which appears in $\Pi_{\text{LLM}}(h_t)$, it reshapes the reward logged at that timestep via
\begin{equation}
\tilde{r}_t = r_t + \beta \cdot w(a_t)
\label{eq:rs}
\end{equation}

where $\beta>0$ is a global scalar bonus parameter and $w(a_t)\in(0,1]$ is the weight of the tier matched by $a_t$. We fix $\beta = 0.01$ for all experiments, so $\tilde{r}_t$ always satisfies $0\leq\tilde{r}_t-r_t\leq\beta$. Note that the LLM is never invoked during action selection. It can only influence training through reward shaping given our constrained host-level action interface: unparsable recommendations are discarded, and PPO solely determines which action will be executed. If PPO chooses an action which is not matched by any tier, then $\tilde{r}_t=r_t$. Otherwise, the original environment reward is kept but can be supplemented by up to $\beta$. Thus, an erroneous recommendation cannot overwrite PPO's selected action directly nor add more than $\beta$ to a single reward. Reward shaping also only happens at logged spreading-state timesteps, and host policies eventually expire after some fixed number of batches and revert to weak defaults.

\subsection{Policy Optimization and Advice Merging}

The RL agent maintains standard policy $\pi_\theta(\cdot | o_t)$ and value network $V_\psi(o_t)$. Given a trajectory buffer $\mathcal{B}$, the agent computes generalized advantage estimates $\hat{A}_t$ using the shaped rewards $\tilde{r}_t$ and updates its parameters with a clipped surrogate policy gradient objective~\cite{schulman2017proximal} at every iteration.

To maximize the value of infrequent LLM consultations, newly generated host policies are \emph{merged} into a global advice dictionary rather than replacing earlier entries. The current guidance mechanism also applies a conservative fallback when a spreading host in the current rollout is not explicitly mentioned by the latest validated LLM response: that host is refreshed with a full-weight \texttt{Restore} tier. In our method, stale host policies expire after a fixed number of batches and revert to the weak default priors, ensuring the active advice dictionary remains focused on the current failure frontier. The complete training procedure is outlined in Algorithm~\ref{alg:llm_ppo}.

\begin{algorithm}[ht]
\caption{Ask the Expert: LLM-Guided RL via Reward Shaping}
\label{alg:llm_ppo}
\textbf{Require:} Env $\mathcal{E}$, policy $\pi_\theta$, value net $V_\psi$, warmup $W_{\mathrm{warm}}$, interval $K_{\mathrm{update}}$, bag size $B_{\max}$, shaping parameter $\beta$
\begin{algorithmic}[1]
\State Init $\Pi_{\text{LLM}}\!\leftarrow\!\text{Weak Default Tiers}$, $\mathcal{B}\!\leftarrow\!\emptyset$, $\mathcal{C}\!\leftarrow\!\emptyset$, $\mathcal{H}\!\leftarrow\!\emptyset$
\For{iteration $n=1,2,\dots$}
    \For{$t=0$ \textbf{to} $T-1$}
        \State Observe $o_t$; $a_t\!\sim\!\pi_\theta(\cdot|o_t)$
        \State Execute $a_t$; observe $r_t$, $o_{t+1}$
        \State Encode reward-breakdown telemetry via $\phi$ into compromised, spreading, and clean host sets
        \State Construct semantic summary $\zeta_t$ from the encoded host sets
        \State $S_t \leftarrow$ spreading hosts recovered by $\phi$
        \State Store $(o_t, a_t, r_t)$ in $\mathcal{B}$
        \If{$S_t \ne \emptyset$}
            \State $h_t \leftarrow$ first host in $S_t$
            \State Record $(t,h_t)$ as a hard-state position for this rollout
            \State Add $(\zeta_t,a_t,r_t,\Delta r_t,n)$ to $\mathcal{C}$
        \EndIf
    \EndFor
    \State Mark outcome of previous LLM call in $\mathcal{H}$ using this rollout's mean reward
    \State Expire stale host policies in $\Pi_{\text{LLM}}$ back to weak defaults
    \For{each recorded hard-state position $(t,h_t)$ in the rollout}
        \If{$a_t^{\mathcal{B}} \in \Pi_{\text{LLM}}(h_t)$}
            \State $\tilde{r}_t \leftarrow r_t + \beta \cdot w(a_t^{\mathcal{B}})$ in $\mathcal{B}$
        \EndIf
    \EndFor
    \If{$n \geq W_{\mathrm{warm}}$ \textbf{and} $(n-W_{\mathrm{warm}})\bmod K_{\mathrm{update}}=0$ \textbf{and} $|\mathcal{C}| > 0$}
        \State $\mathrm{entries} \leftarrow \mathrm{SampleRecentFrontier}(\mathcal{C}, n_s)$
        \State $p \leftarrow \mathrm{BuildPrompt}(\mathrm{entries}, \mathcal{H})$
        \State $\mathcal{M}\!\leftarrow\!\mathrm{Validate}(\mathrm{LLM}(p))$
        \State $\Pi_{\text{LLM}} \leftarrow \text{Merge}(\Pi_{\text{LLM}}, \text{ExpandTiered}(\mathcal{M}))$
        \State For spreading hosts from this rollout not addressed by $\mathcal{M}$, set $\Pi_{\text{LLM}}(h) \leftarrow \text{full-weight \texttt{Restore} tier}$
        \State Append validated suggestions to $\mathcal{H}$
        \label{line:llm}
    \EndIf
    \State Compute $\hat{A}_t$ via GAE on the shaped rollout rewards; update $\pi_\theta$, $V_\psi$
    \State $\mathcal{B} \leftarrow \emptyset$
\EndFor
\end{algorithmic}
\end{algorithm}

\section{Experimental Evaluation}
\label{sec:experiments}

\subsection{Experimental Details}

We evaluate on two TTCP CAGE Challenges~\cite{kiely2023autonomous} built on the CybORG simulation environment~\cite{cyborg2021}: CAGE Challenge~1 (CC1) and Challenge~2 (CC2). Both environments model a medium-sized enterprise network with 13 hosts across user, enterprise, and operational subnets, with episodes of 100 timesteps and partial defender observations from host alerts and session telemetry. CC1 provides a 54-action discrete space comprising two environment-wide actions (Sleep, Monitor) and four host-targeted action types (Analyse, Remove, Restore, Misinform) applied across 13 hosts, where Misinform deploys decoy services intended to delay Red and expose attacker activity. CC2 provides a larger 145-action space in which Misinform is replaced by service-specific Decoy actions. All results are evaluated against \texttt{B\_lineAgent} and \texttt{RedMeanderAgent}~\cite{kiely2023autonomous}.

All reported runs used Stable-Baselines3 PPO~\cite{sb3} with an MLP policy, learning rate $3\times 10^{-4}$, $n_\mathrm{steps}=2048$, batch size 128, 6 PPO epochs per rollout, $\gamma=0.99$, clip range $0.2$, and entropy coefficient $0.01$, with synchronized train/evaluation normalization. 
As a reward-shaping control, we additionally evaluate an observation-based PBRS baseline~\cite{ng1999_rewardshaping}, $r_t' = r_t + \lambda[\gamma\Phi(o_{t+1})-\Phi(o_t)]$, for $\lambda\in\{0.1,0.3,1.0\}$, where $\Phi(o_t)$ is the negative defender-visible threat score, constructed from compromise severity and host criticality such that more severe compromise of higher-criticality hosts yields lower potential.
For both attacker settings, the shared \emph{Ask the Expert} mechanism used $B_{\max}=200$, $n_s=20$, $W_{\mathrm{warm}}=20$ rollouts, a cadence of 10 rollouts per LLM query, and a fixed shaping bonus $\beta=0.01$ with Gemini 3.1 Flash-Lite. Default host priors were weak \texttt{Restore} and \texttt{Remove} tiers with weights $(0.1, 0.05)$; post-LLM host policies used weights $(1.0, 0.5, 0.3, 0.15)$ for the primary, restore, remove, and analyse tiers, respectively. Prompt construction used a recent-failure frontier with host diversity capping, as well as a host-policy time-to-live after which stale advice reverted to defaults. In CC2, PPO operated in the full 145-action space and the LLM was restricted to a host-level \texttt{Analyse}/\texttt{Remove}/\texttt{Restore} remediation vocabulary; in CC1, PPO operated in the 54-action space and the LLM vocabulary additionally included \texttt{Misinform}. In both cases only the prompt doctrine was specialized to attacker structure. All reported evaluations were deterministic over 20 episodes.

\subsection{Performance Results}

\begin{figure}[ht]
\centering
\includegraphics[width=0.975\columnwidth]{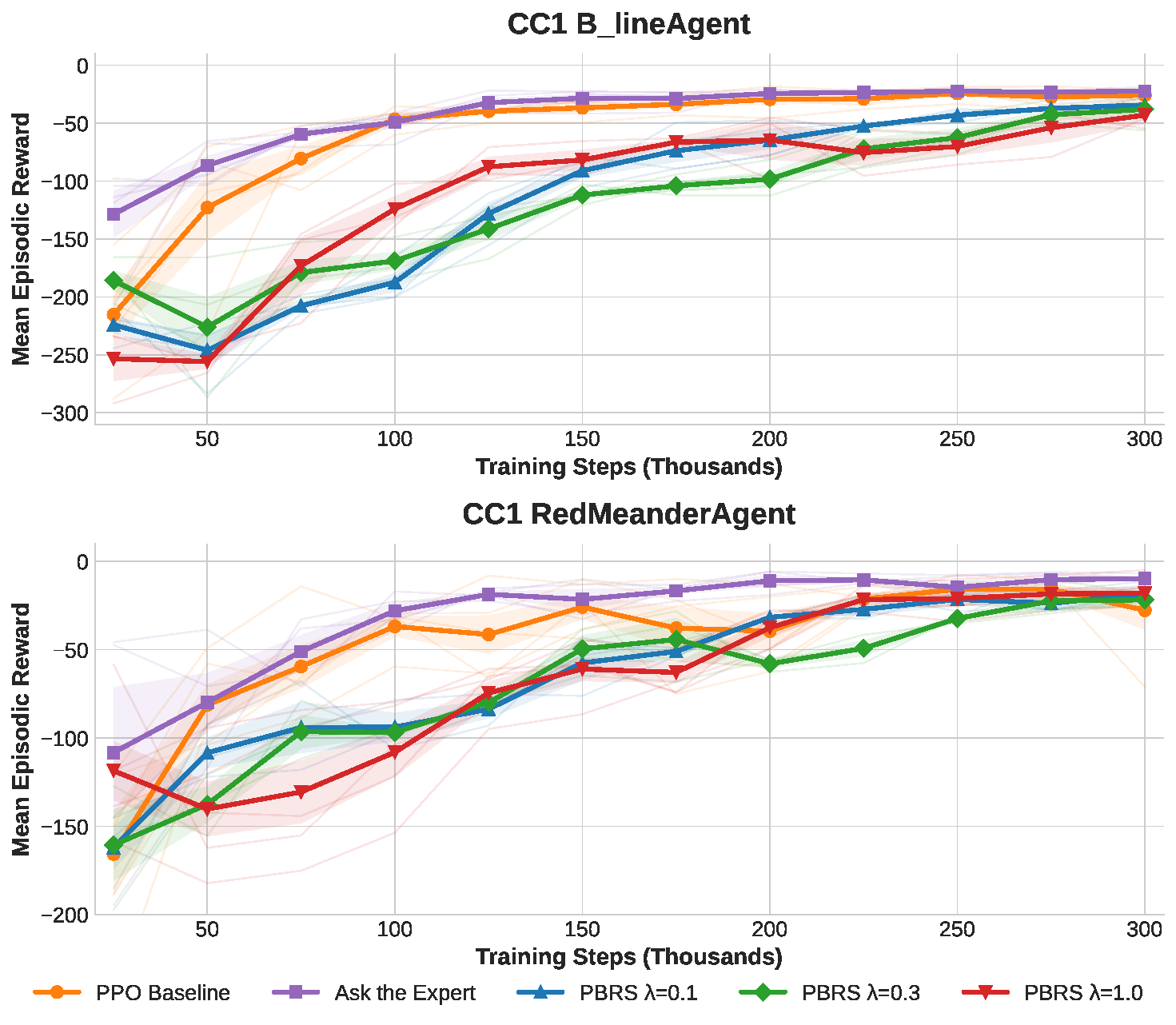}
\caption{Mean checkpoint evaluation reward for PPO, \emph{Ask the Expert}, and PBRS variants in CC1. Top: \texttt{B\_lineAgent}. Bottom: \texttt{RedMeanderAgent}. Shaded bands denote $\pm 1$ standard error across independent training seeds.}
\label{fig:cc1_results}
\end{figure}

Shown in Figure~\ref{fig:cc1_results} are the CC1 results. Against \texttt{B\_lineAgent}, \emph{Ask the Expert} improves early learning over PPO, reaching $-86.7$ vs.\ $-122.9$ at 50k steps and $-59.6$ vs.\ $-80.5$ at 75k. It also outperforms all evaluated PBRS variants at every reported checkpoint and finishes at $-22.3$, compared with $-25.9$ for PPO and $-34.4$ for the best PBRS mean at 300k. Against \texttt{RedMeanderAgent}, the method reaches $-27.9$ at 100k compared with $-36.8$ for PPO and $-93.8$ for the best PBRS mean at that checkpoint, and retains the best terminal mean ($-9.7$ vs.\ $-27.7$ for PPO and $-18.0$ for PBRS).

\begin{figure}[ht]
\centering
\includegraphics[width=0.975\columnwidth]{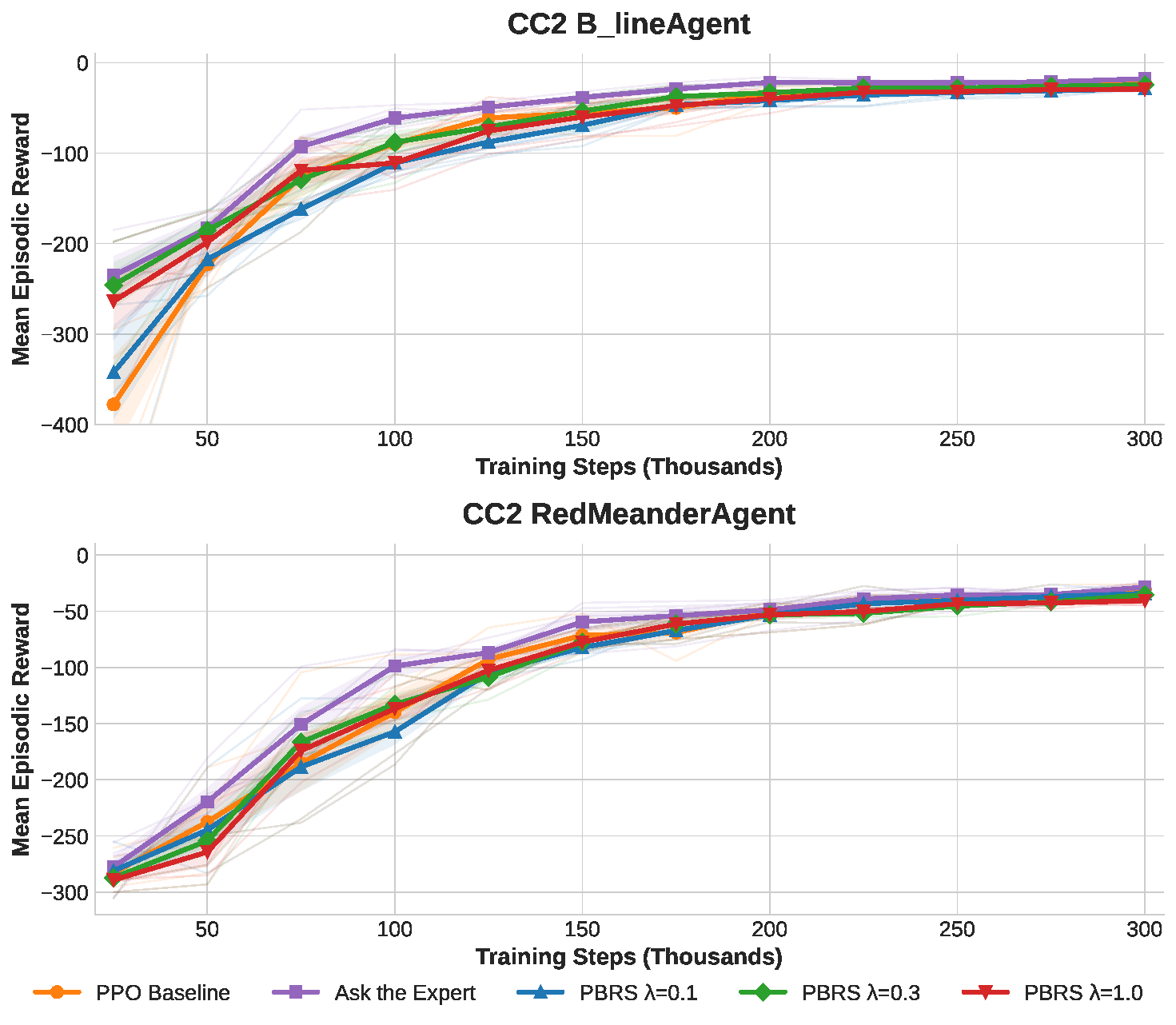}
\caption{Mean checkpoint evaluation reward for PPO, \emph{Ask the Expert}, and PBRS variants in CC2. Top: \texttt{B\_lineAgent}. Bottom: \texttt{RedMeanderAgent}. Shaded bands denote $\pm 1$ standard error across independent training seeds.}
\label{fig:cc2_results}
\end{figure}

Figure~\ref{fig:cc2_results} shows the same pattern on CC2. Against \texttt{B\_lineAgent}, \emph{Ask the Expert} reaches $-61.3$ at 100k versus $-88.1$ for the best PBRS mean at that checkpoint and $-89.7$ for PPO, and finishes at $-17.9$ compared with $-24.7$ and $-21.5$, respectively. Against \texttt{RedMeanderAgent}, the method reaches $-98.8$ at 100k versus $-132.7$ for the best PBRS mean at that checkpoint and $-139.9$ for PPO; the gap narrows later, but \emph{Ask the Expert} retains the best mean at 300k ($-28.6$ vs.\ $-34.2$ for PBRS and $-30.4$ for PPO). In all four settings, \emph{Ask the Expert} achieves higher mean reward than every evaluated PBRS variant throughout the reported mean learning curves.

\subsection{Mechanistic Analysis of Frontier Sampling}

\begin{figure}[!t]
\centering
\includegraphics[width=0.95\columnwidth]{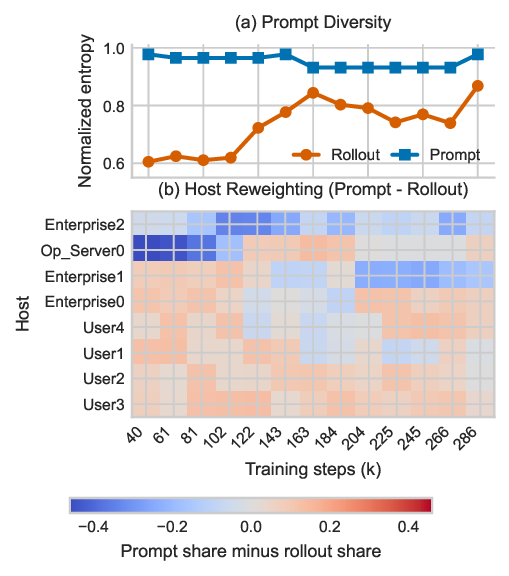}
\vspace{-10pt}
\caption{Frontier-sampling mechanics from one representative \texttt{B\_lineAgent} run. Panel (a) compares the normalized Shannon entropy~\cite{shannon1948_communication} of host frequencies in the raw rollout stream and in the sampled prompt over training. Panel (b) shows the host-wise reweighting induced by the sampler, measured as prompt share minus rollout share at each query. Positive values indicate that a host is emphasized more in the prompt than in the rollout stream.}
\label{fig:frontier_mechanics}
\end{figure}

Figure~\ref{fig:frontier_mechanics} shows another perspective on how the recent-frontier sampler behaves on one representative \texttt{B\_lineAgent} run. Panel (a) shows that the sampled prompt remains consistently more diverse than the raw rollout stream over the course of training. In other words, the sampler is exposing the LLM to a wider variety of hosts than the raw rollout frequencies would suggest. Panel (b) explains how this happens. Early in training, heavily represented hosts such as \texttt{Op\_Server0} and \texttt{Enterprise2} are deemphasized relative to their rollout frequency, while other hosts receive more prompt mass; later, the emphasis shifts as the active frontier moves. This figure is not intended as a separate performance result, but to illustrate that frontier sampling keeps successive LLM queries focused on a wider set of frontier-relevant states instead of repeatedly mirroring the most common host in the raw rollout stream.

\section{Discussion}
\label{sec:discussion}
Across CC1 and CC2, \emph{Ask the Expert} attains higher mean reward than every evaluated PBRS variant at each reported checkpoint for both attacker types, while also improving substantially over PPO during early and intermediate training. The separation is especially pronounced in CC1 and against \texttt{RedMeanderAgent}; in CC2, the PBRS gap narrows as training progresses, but \emph{Ask the Expert} retains the strongest terminal mean in both attacker settings. These results show that the observed gains are not reproduced by the evaluated observation-based potential shaping baselines, while deployment remains fully policy-based.

The frontier-sampling diagnostic (Figure~\ref{fig:frontier_mechanics}) validates that prompt construction continues to expose the LLM to diverse frontier-relevant states rather than collapsing onto the most frequent host in the raw rollout stream, confirming that the diversity benefit of the sampler persists throughout training.

\emph{Ask the Expert}'s results set it as a viable asymmetric-training approach for self-defense systems: semantic reasoning is provided only where it can do the most good, at training time, while deployment remains lightweight and fully policy-based.

\section{Conclusion}
\label{sec:conclusion}

We introduced \emph{Ask the Expert}, a training-time LLM-guided reward-shaping framework for autonomous cyber defense. The framework converts defender-visible compromise signals into semantic summaries, intermittently queries an LLM, and distills constrained host-level advice into tiered reward bonuses while leaving deployment as a pure RL policy. On CAGE CC1 and CC2 under two attacker types, the method improves sample efficiency over PPO and achieves higher mean reward than every evaluated PBRS variant throughout the reported learning curves. Although the gap to PBRS narrows during late training in some settings, \emph{Ask the Expert} retains the strongest terminal mean in all four settings.

\bibliographystyle{IEEEtran}
\bibliography{references}

\begin{thebibliography}{10}
\providecommand{\url}[1]{#1}
\csname url@samestyle\endcsname
\providecommand{\newblock}{\relax}
\providecommand{\bibinfo}[2]{#2}
\providecommand{\BIBentrySTDinterwordspacing}{\spaceskip=0pt\relax}
\providecommand{\BIBentryALTinterwordstretchfactor}{4}
\providecommand{\BIBentryALTinterwordspacing}{\spaceskip=\fontdimen2\font plus
\BIBentryALTinterwordstretchfactor\fontdimen3\font minus \fontdimen4\font\relax}
\providecommand{\BIBforeignlanguage}[2]{{%
\expandafter\ifx\csname l@#1\endcsname\relax
\typeout{** WARNING: IEEEtran.bst: No hyphenation pattern has been}%
\typeout{** loaded for the language `#1'. Using the pattern for}%
\typeout{** the default language instead.}%
\else
\language=\csname l@#1\endcsname
\fi
#2}}
\providecommand{\BIBdecl}{\relax}
\BIBdecl

\bibitem{smiliotopoulos2024detecting}
C.~Smiliotopoulos, G.~Kambourakis, and C.~Kolias, ``Detecting lateral movement: A systematic survey,'' \emph{Heliyon}, vol.~10, no.~4, 2024.

\bibitem{schulman2017proximal}
J.~Schulman, F.~Wolski, P.~Dhariwal, A.~Radford, and O.~Klimov, ``Proximal policy optimization algorithms,'' \emph{arXiv preprint arXiv:1707.06347}, 2017.

\bibitem{nguyen2019deep}
T.~T. Nguyen and V.~J. Reddi, ``Deep reinforcement learning for cyber security,'' \emph{IEEE Transactions on Neural Networks and Learning Systems}, vol.~34, no.~8, pp. 3779--3795, 2021.

\bibitem{hu2021survey}
W.~Hu, X.~Liao \emph{et~al.}, ``A survey on reinforcement learning for cyber security,'' \emph{IEEE Communications Surveys \& Tutorials}, vol.~23, no.~2, pp. 1353--1381, 2021.

\bibitem{bates2023reward}
E.~Bates, V.~Mavroudis, and C.~Hicks, ``Reward shaping for happier autonomous cyber security agents,'' in \emph{Proceedings of the 16th ACM Workshop on Artificial Intelligence and Security}, 2023, pp. 221--232.

\bibitem{palmer2023deep}
G.~Palmer, C.~Parry, D.~J. Harrold, and C.~Willis, ``Deep reinforcement learning for autonomous cyber defence: A survey,'' \emph{arXiv preprint arXiv:2310.07745}, 2023.

\bibitem{ng1999_rewardshaping}
A.~Y. Ng, D.~Harada, and S.~Russell, ``Policy invariance under reward transformations: Theory and application to reward shaping,'' in \emph{Proceedings of the Sixteenth International Conference on Machine Learning}, 1999, pp. 278--287.

\bibitem{xu2025llm4sec}
\BIBentryALTinterwordspacing
H.~Xu, S.~Wang, N.~Li, K.~Wang, Y.~Zhao, K.~Chen, T.~Yu, Y.~Liu, and H.~Wang, ``Large language models for cyber security: A systematic literature review,'' \emph{ACM Trans. Softw. Eng. Methodol.}, Sep. 2025, just Accepted. [Online]. Available: \url{https://doi.org/10.1145/3769676}
\BIBentrySTDinterwordspacing

\bibitem{purves2024causally}
T.~Purves, K.~G. Kyriakopoulos, S.~Jenkins, I.~Phillips, and T.~Dudman, ``Causally aware reinforcement learning agents for autonomous cyber defence,'' \emph{Knowledge-Based Systems}, vol. 304, p. 112521, 2024.

\bibitem{zhang2023large}
Y.~Zhang \emph{et~al.}, ``Large language models for cyber security: A systematic review,'' \emph{IEEE Security \& Privacy}, 2023.

\bibitem{ma2024eureka}
\BIBentryALTinterwordspacing
Y.~J. Ma, W.~Liang, G.~Wang, D.-A. Huang, O.~Bastani, D.~Jayaraman, Y.~Zhu, L.~Fan, and A.~Anandkumar, ``Eureka: Human-level reward design via coding large language models,'' in \emph{The Twelfth International Conference on Learning Representations}, 2024. [Online]. Available: \url{https://openreview.net/forum?id=IEduRUO55F}
\BIBentrySTDinterwordspacing

\bibitem{pmlr-v202-du23f}
\BIBentryALTinterwordspacing
Y.~Du, O.~Watkins, Z.~Wang, C.~Colas, T.~Darrell, P.~Abbeel, A.~Gupta, and J.~Andreas, ``Guiding pretraining in reinforcement learning with large language models,'' in \emph{Proceedings of the 40th International Conference on Machine Learning}, ser. Proceedings of Machine Learning Research, A.~Krause, E.~Brunskill, K.~Cho, B.~Engelhardt, S.~Sabato, and J.~Scarlett, Eds., vol. 202.\hskip 1em plus 0.5em minus 0.4em\relax PMLR, 23--29 Jul 2023, pp. 8657--8677. [Online]. Available: \url{https://proceedings.mlr.press/v202/du23f.html}
\BIBentrySTDinterwordspacing

\bibitem{martinez2026stackelberg}
F.~Martinez, T.~Li, Y.~Lu, and J.~Chen, ``Stackelberg coupling of online representation learning and reinforcement learning,'' in \emph{The Fourteenth International Conference on Learning Representations}, 2026.

\bibitem{martinez2025context}
F.~Martinez-Lopez, T.~Li, Y.~Lu, and J.~Chen, ``In-context reinforcement learning via communicative world models,'' \emph{arXiv preprint arXiv:2508.06659}, 2025.

\bibitem{kim-tao25col}
K.~Hammar, T.~Li, R.~Stadler, and Q.~Zhu, ``Adaptive security response strategies through conjectural online learning,'' \emph{{IEEE} Transactions on Information Forensics and Security}, vol.~20, pp. 4055--4070, 2025.

\bibitem{cyborg2021}
J.~Standen \emph{et~al.}, ``Cyborg: A cyber operations research gym for autonomous agents,'' \emph{Proceedings of the AAAI Conference on Artificial Intelligence}, 2021.

\bibitem{schaul2016_per}
\BIBentryALTinterwordspacing
T.~Schaul, J.~Quan, I.~Antonoglou, and D.~Silver, ``Prioritized experience replay,'' in \emph{International Conference on Learning Representations}, 2016. [Online]. Available: \url{https://dblp.org/rec/journals/corr/SchaulQAS15}
\BIBentrySTDinterwordspacing

\bibitem{griffith2013_policyshaping}
\BIBentryALTinterwordspacing
S.~Griffith, K.~Subramanian, J.~Scholz, C.~L. Isbell, and A.~L. Thomaz, ``Policy shaping: Integrating human feedback with reinforcement learning,'' in \emph{Advances in Neural Information Processing Systems}, vol.~26, 2013. [Online]. Available: \url{http://hdl.handle.net/1853/53270}
\BIBentrySTDinterwordspacing

\bibitem{wang2026collaborative}
\BIBentryALTinterwordspacing
Y.~Wang, X.~Wu, J.~Farooq, T.~Li, and J.~Chen, ``Collaborative cloud-edge computing via llm-guided constrained reinforcement learning,'' \emph{TechRxiv}, 2026. [Online]. Available: \url{https://www.techrxiv.org/doi/full/10.36227/techrxiv.177155994.44918685/v1}
\BIBentrySTDinterwordspacing

\bibitem{xie2024textreward}
\BIBentryALTinterwordspacing
T.~Xie, S.~Zhao, C.~H. Wu, Y.~Liu, Q.~Luo, V.~Zhong, Y.~Yang, and T.~Yu, ``Text2reward: Reward shaping with language models for reinforcement learning,'' in \emph{The Twelfth International Conference on Learning Representations}, 2024. [Online]. Available: \url{https://openreview.net/forum?id=tUM39YTRxH}
\BIBentrySTDinterwordspacing

\bibitem{zhang2025progress}
\BIBentryALTinterwordspacing
X.~Zhang, N.~Gao, X.~Jiang, Y.~Chen, Y.~Pan, M.~Zhang, and Y.~Deng, ``Progress reward model for reinforcement learning via large language models,'' in \emph{The Thirty-ninth Annual Conference on Neural Information Processing Systems}, 2025. [Online]. Available: \url{https://openreview.net/forum?id=TJhHb6CscW}
\BIBentrySTDinterwordspacing

\bibitem{sun2025_card}
\BIBentryALTinterwordspacing
S.~Sun, R.~Liu, J.~Lyu, J.-W. Yang, L.~Zhang, and X.~Li, ``A large language model-driven reward design framework via dynamic feedback for reinforcement learning,'' \emph{Knowledge-Based Systems}, 2025. [Online]. Available: \url{https://www.sciencedirect.com/science/article/pii/S0950705125011104}
\BIBentrySTDinterwordspacing

\bibitem{li2026_llmhrl}
\BIBentryALTinterwordspacing
Q.~Li, B.~J. Pang, Y.~Song, H.~Fu, Q.~Xu, X.~Yuan, X.~Xu, and C.~Zhang, ``Large language model assisted hierarchical reinforcement learning training,'' \emph{Information Sciences}, 2026. [Online]. Available: \url{https://colab.ws/articles/10.1016%2Fj.ins.2025.122688}
\BIBentrySTDinterwordspacing

\bibitem{homoliak2019insight}
\BIBentryALTinterwordspacing
I.~Homoliak, F.~Toffalini, J.~Guarnizo, Y.~Elovici, and M.~Ochoa, ``Insight into insiders and it: A survey of insider threat taxonomies, analysis, modeling, and countermeasures,'' \emph{ACM Computing Surveys}, vol.~52, no.~2, p. 1–40, Apr. 2019. [Online]. Available: \url{http://dx.doi.org/10.1145/3303771}
\BIBentrySTDinterwordspacing

\bibitem{kiely2023autonomous}
M.~Kiely, D.~Bowman, M.~Standen, and C.~Moir, ``On autonomous agents in a cyber defence environment,'' \emph{arXiv preprint arXiv:2309.07388}, 2023.

\bibitem{sb3}
\BIBentryALTinterwordspacing
A.~Raffin, A.~Hill, A.~Gleave, A.~Kanervisto, M.~Ernestus, and N.~Dormann, ``Stable-baselines3: Reliable reinforcement learning implementations,'' \emph{Journal of Machine Learning Research}, vol.~22, no. 268, pp. 1--8, 2021. [Online]. Available: \url{http://jmlr.org/papers/v22/20-1364.html}
\BIBentrySTDinterwordspacing

\bibitem{shannon1948_communication}
C.~E. Shannon, ``A mathematical theory of communication,'' \emph{Bell System Technical Journal}, vol.~27, no. 3--4, pp. 379--423, 623--656, 1948.

\end{thebibliography}
\end{document}